\documentclass[a4paper,11pt]{article}
\pdfoutput=1 % if your are submitting a pdflatex (i.e. if you have
\usepackage{jheppub} % for details on the use of the package, please
\usepackage[T1]{fontenc} % if needed
\pdfoutput=1

\usepackage{graphicx}
\usepackage{epsfig}
\usepackage{empheq}
\usepackage{mathbbol}
\usepackage{wasysym}
\usepackage{bbm}
\usepackage{bm}
\usepackage{color}
\usepackage{amssymb,amsmath}
\usepackage{epstopdf}
\usepackage{float}
\usepackage{color}
\usepackage{subfig}
\graphicspath{{./png/}}

\def\cO{{\mathcal O}}

\newcommand{\be}{\begin{equation}}
\newcommand{\ee}{\end{equation}}
\newcommand{\bea}{\begin{eqnarray}}
\newcommand{\eea}{\end{eqnarray}}

\title{Neutrino Mixing and Masses from a Minimum Principle}
\author[a,b]{R. Alonso,} \author[a,b]{M. B. Gavela,} \author[b,c]{G. Isidori,} \author[b,d]{L. Maiani }
\affiliation[a]{ Instituto de F\'{\i}sica Te\'orica UAM/CSIC and Departamento de F\'isica Te\'orica\\
 Universidad Aut\'onoma de Madrid,~Cantoblanco,~28049 Madrid, Spain}
\affiliation[b]{CERN, PH-TH, 1211 Gen\`eve 23, Switzerland}
\affiliation[c]{ INFN Laboratori Nazionali di Frascati, Via E. Fermi 40, Frascati, I-00044, Italy}
\affiliation[d]{ Dipartimento di Fisica, Universit\`a di Roma "La Sapienza",\\ Piazzale A Moro 5, Roma, I-00185, Italy}

\abstract{
We analyze the structure of  quark and  lepton  mass matrices under the hypothesis that they are 
determined from a minimum principle applied to a generic potential invariant under the $\left[SU(3)\right]^5\otimes  {\mathcal O}(3)$ flavor symmetry, 
acting on Standard Model fermions and right-handed neutrinos. Unlike the quark case, we show that hierarchical masses for charged leptons are naturally accompanied 
by degenerate Majorana neutrinos with one {mixing} angle close to maximal, a second potentially large, a third one necessarily small, 
{and one maximal relative Majorana phase.}
Adding small perturbations the predicted structure for the neutrino mass matrix is in excellent agreement with present observations and 
could be tested in the near future via neutrino-less double beta decay and cosmological measurements. The generalization of these results to arbitrary sew-saw models is also discussed.
}
\begin{document}

\maketitle
\flushbottom

\section{Introduction} 
The gauge interactions of the Standard Model (SM) admit a large, global, flavor symmetry. Matter fields in the SM are described by quark and lepton doublets, $q_L$ and $\ell_L$, and by right-handed  singlets corresponding to $up$ and $down$ quarks and to electron-like leptons: $U_R$, $D_R$, $E_R$.  With three quark and lepton generations, massless neutrinos, and omitting $U(1)$ factors unessential to the present work, 
the flavor group is~\cite{Chivukula:1987py}: 

\be 
{\cal G}_0=\left[SU(3)\right]^5=SU(3)_{q}\otimes SU(3)_{U}\otimes SU(3)_{D}\otimes SU(3)_{{\ell}}\otimes SU(3)_{E}\,.
\label{gnumassless}
\ee

Masses for the observed neutrinos can be generated with the see-saw mechanism~\cite{Minkowski:1977sc}, by introducing at least two generations of additional Majorana neutrinos, $N_i$. The latter are endowed with a Majorana mass matrix with possibly  
large eigenvalues, and coupled to the lepton doublets by Yukawa interactions.
In analogy with the quark sector, here we assume three Majorana generations. We also assume the maximal flavor symmetry acting on the $N_i$ in the limit of vanishing Yukawa couplings but  non-vanishing Majorana masses, i.e. ${\cal O}(3)$. The flavor group for this case is~\cite{Cirigliano:2005ck}:
\be
{\cal G}=\left[SU(3)\right]^5\otimes  {\mathcal O}(3)\,.
\label{gnumass}
\ee

 The large flavor group in Eq.~(\ref{gnumass}), of course, does not correspond to observed symmetries. 
 In the SM, global symmetries are explicitly broken by the Yukawa couplings of matter fields to the $SU(2)_L$ scalar doublet,
whose vacuum expectation value (vev) is  responsible for the
 spontaneous breaking of the $SU(2)_L\otimes U(1)$ gauge symmetry~\cite{Englert:1964et}.
 
  Explicit breaking avoids the presence of unseen Goldstone or pseudo Goldstone bosons with the disadvantage, however, that values and textures of the Yukawa coupling matrices are out of reach of any theoretical consideration.  
 Also, this picture presents the danger that any new physics will bring in new independent amplitudes to Flavor Changing Neutral Current  (FCNC)
 processes, spoiling the extremely good SM predictions, see e.g.~Ref.~\cite{Isidori:2010kg}. This can be avoided by the so called Minimal Flavor Violation (MFV) principle~\cite{D'Ambrosio:2002ex},  the statement that in any new physics sector the only sources of flavor symmetry breaking accessible at low energies are {\it the same} Yukawa couplings.\footnote{See Ref.~\cite{Cirigliano:2005ck,Davidson:2006bd,Gavela:2009cd,Alonso:2011jd,Joshipura:2009gi} for a discussion about the implementation of the MFV principle in models with non-vanishing neutrino masses.} Even the  MFV principle, however, does not provide an explanation of the observed pattern of  masses and mixings of quarks and leptons.

In this paper we want to discuss a different option, namely that the Yukawa couplings  are the vacuum expectation values of {\it Yukawa fields}, to be determined by  a minimum principle applied to some potential, $V({\it Y})$, invariant under the full flavor group ${\cal G}$.   In this case, one may use group theoretical methods to identify the {\it natural extrema} and characterize the texture of the resulting Yukawa matrices.

 The simplest realization of the idea of a dynamical character for the Yukawa couplings is to assume that
 \be
 {\it Y}=\frac{\langle 0|\Phi|0\rangle }{\Lambda}
 \label{vev1}
 \ee
 with $\Lambda$ some high energy scale and $\Phi$ a set of scalar fields with transformation properties such as to make invariant the effective Lagrangians and the potential $V({\it Y})$ under $\cal G$.  
 To avoid the problem of unseen Goldstone bosons,  $ {\cal G}$ may be in fact a local gauge symmetry broken at the scale $\Lambda$, with an appropriate Higgs mechanism, see e.g.~Ref.~\cite{Grinstein:2010ve}.

The idea that quark masses could arise from the minimum of a chiral $SU(3)\otimes SU(3)$ symmetric potential was considered in the sixties by N. Cabibbo, in the attempt to determine theoretically the value of the Cabibbo angle, and group theoretical  methods  were established in Refs.~\cite{Michel:1970mua} and~\cite{Cabibbo:1970rza} to identify the natural extrema of the potential. Further attempts towards a dynamical origin of the Yukawa couplings, employing either $G_0$ or other flavor groups and different flavor-breaking fields,  have been proposed in Refs.~\cite{Froggatt:1978nt,Anselm:1996jm, Barbieri:1999km,Berezhiani:2001mh,Harrison:2005dj,Feldmann:2009dc,Alonso:2011yg,Nardi:2011st,Alonso:2012fy,Espinosa:2012uu,Alonso:2013mca}. In all cases, the analysis reduces to determining the possible invariants made out of Yukawa fields, and explore their extrema.
  
 In this paper, we apply the theoretical considerations devised in Refs.~\cite{Michel:1970mua,Cabibbo:1970rza} for chiral symmetry, to the flavor group ${\cal G}$ in Eq.~(\ref{gnumass}). Interestingly,  we find different textures for the Yukawa matrices of quarks and leptons.
For quarks, we find the hierarchical mass pattern of the third {\it vs} the first two generations and unity Cabibbo-Kobayashi-Maskawa (CKM) matrix. 
 However, for leptons, hierarchical masses for charged leptons are accompanied by degenerate Majorana neutrinos with one, potentially two, large mixing angles and {one maximal relative Majorana phase}. 
 
Both textures are close to the real situation. Moreover, if we add small perturbations to the neutrino mass matrix, we obtain a realistic pattern of mass differences and a Pontecorvo-Maki-Nakagawa-Sakata (PNMS) mixing matrix close to the bimaximal or tribimaximal mixing with small $\theta_{13}$, without having to resort to symmetries under discrete groups~\cite{Harrison:2002er}.
 The prediction that large mixing angles are correlated to Majorana degenerate neutrinos with an average mass that we estimate could be as large as
\be
m_\nu \approx  0.1~{\rm eV},
\ee
could be tested in a not too distant future in $0\nu\nu$ double beta decay, see e.g.~Ref.~\cite{pdg}, and possibly in cosmological measurements~\cite{Ade:2013zuv,Ade:2013lmv}.

As anticipated, the idea that the pattern of the Yukawa couplings can be derived by the vev of appropriate fields is not new. 
In the present context, it was introduced by Froggat and Nielsen~\cite{Froggatt:1978nt} for  a global flavor $U(1)$ symmetry.  It was retaken for quarks 
in Refs.~\cite{Anselm:1996jm,Berezhiani:2001mh,Alonso:2011yg,Nardi:2011st,Espinosa:2012uu,Alonso:2012fy}, 
employing the symmetry group ${\cal G}_0$  and analyzing the allowed minima of renormalizable potentials. The case of leptons with Majorana neutrinos was first analysed in a particular model~\cite{Alonso:2012fy}, and later in general~\cite{Alonso:2013mca}, in the context of the type I see-saw model 
 with two heavy degenerate right-handed (RH) neutrinos, corresponding to an $\cO(2)$ flavor symmetry, for both two and three light generations. In
 Ref.~\cite{Alonso:2012fy,Alonso:2013mca} it has been shown that  the minimum of the scalar potential  does allow a maximal mixing 
angle  --in contrast to the quark case-- and a maximal Majorana phase, associated to degenerate neutrinos.  
For three RH neutrinos, another two sizable angles were obtained within $\cO(2)$, albeit at the price of adding  a supplementary flavor vector component to the Yukawa field. The $\cO(3)$ flavor symmetry for light neutrinos had  been  considered from a different perspective
in Ref.~\cite{Blankenburg:2012nx}, where a large angle is obtained after the introduction of small perturbations to the symmetric situation and almost degenerate neutrinos are envisaged.
Our results provide a generalization of previous findings to the case of three heavy RH neutrinos. Most important, we show that the 
configuration with degenerate light neutrinos and  one maximal mixing angle and Majorana phase holds irrespectively from the renormalizability of the potential and 
is therefore stable under radiative corrections.

\section{ Couplings and definitions} 
\label{coupl}
Yukawa couplings can be organised in a set of matrices, ${\it Y}$, which appear in the flavor symmetry breaking, gauge symmetry conserving, lagrangian:
 \be
- {\cal L}_Y={\bar q}_L {\it Y}_D H D_R +{\bar q}_L {\it Y}_U {\tilde H} U_R+{\bar \ell}_L {\it Y}_E H E_R+{\bar \ell}_L {\it Y}_\nu {\tilde H}  N+ {\rm h.c.}+\frac{M}{2}N
\gamma_0 N~.
\label{yukawa}
\ee
 Majorana representation of gamma matrices is used throughout, $H$ is the scalar doublet and $\tilde H$ its charge conjugate. For quarks and charged leptons, we find:
\bea
&&M_D= v{\it Y}_D~, \qquad M_U=v{\it Y}_U~, \qquad M_E=v{\it Y}_E \label{mass}~,   \qquad v= \langle 0|H|0 \rangle~. 
\label{vev}
\eea
Integrating over the $N$ fields and keeping the light fields  only, one finds, to lowest order :
\be
\frac{M}{2}N\gamma_0 N+({\bar \ell}_L {\it Y}_\nu {\tilde H}  N+ {\rm h.c.}) \quad \longrightarrow  \quad \frac{1}{2}{\bar \ell}_L{\it Y}_\nu {\tilde H} {\tilde H}^T {\it Y}_\nu^T {\ell}^C_L+ {\rm h.c.}
\label{yuseesaw}
\ee
which, upon spontaneous breaking of the gauge symmetry, gives the see-saw formula for the light neutrino mass matrix:
\bea
&&m_\nu=\frac{v^2}{M}  {\it Y}_\nu  {\it Y}_\nu^T~.
\label{numass}
\eea

The ${\cal G}$ transformations on ${\it Y}$, or $\Phi$ in Eq.~(\ref{vev1}), that make the lagrangian invariant are as follows:
\bea
&&Y_U \to U_{q} Y_U U_{U}~, \qquad Y_D \to U_{q} Y_D U_{D}~, \label{quarktransf}\\
&& Y_E \to U_{\ell} Y_E U_{E}~, \qquad Y_\nu \to U_{\ell} Y_\nu {\cal O}^T~,
\label{leptransf}
\eea
with the $U$ unitary and ${\cal O}$ real orthogonal, $3\times 3$ matrices. Using this arbitrariness, we can reduce the Yukawa matrices to a standard form:
\bea
&& Y_U=y_U~, \qquad Y_D=U_{\rm CKM} ~y_D~, \label{standardq}\\
&& Y_E=y_E~, \qquad Y_\nu=U_L ~y_\nu ~\omega~ U_R~,
\label{standardlep}
\eea

 Because the Yukawa matrices are defined in between Fermi fields, see Eq. (\ref{yukawa}), an overall phase can be eliminated by appropriate (flavor-independent) phase redefinitions of the fields $U_R$, $D_R$, $E_R$, and $\ell_L$. Thus we may assume the determinants of $Y_U$, $Y_D$, $Y_E$, and $Y_\nu$ in Eq. (\ref{yukawa})  to be real. In this case, all the $y$ are diagonal, real and positive, all the $U$ are unitary  matrices and $\omega$ is a diagonal phase matrix  of unit determinant.
 For simplicity, we shall assume that the generic fields $\Phi$ in (\ref{vev1}) obey the same (${\cal G}$ invariant) condition of having a real 
 determinant.\footnote{ 
The locations of the extrema determined by the structure of  ${\cal G}$ do not change if we include the extra phases. The latter could be eliminated from the present analysis by the $U(1)$ factors omitted in the definition of  ${\cal G}$, irrespective of the sensitivity of the  anomalous $U(1)$ quark axial current to such field redefinitions. 
 }

In the quark sector we are left with one unitary matrix embodying the CKM mixing.  To read neutrino masses we need in addition to diagonalize the matrix in Eq.~(\ref{numass}). We write:
\bea
&&m_\nu=\frac{v^2}{M}U_L(y_\nu \omega U_R U_R^T \omega y_\nu)U_L^T=U_{\rm PMNS}~\Omega~{\hat m}_\nu~ \Omega ~U_{\rm PMNS}^T~,
\label{numass2}
\eea
which introduces the PMNS mixing matrix, with ${\hat m}_\nu$ diagonal, real and positive, and $\Omega$ a diagonal, Majorana phase matrix.

It follows  from Eq.~(\ref{standardq}) that there are in all 10 independent real parameters (i.e.~invariants) in the quark sector.
To count the real parameters appearing in Eq.~(\ref{standardlep}), we start from neutrinos. We have $4$ parameters in $U_L$, as in the CKM matrix, $3$ real eigenvalues in $y_\nu$ and $3$ parameters in $U_R$ counted as follows: $8$ for a general $3\times 3$ special, unitary matrix, less $3$, corresponding to an orthogonal transformation we may perform on the Majorana fields, less $2$ phases we include in $\omega$. Adding the $3$ real eigenvalues of $Y_E$, we obtain a total of $15$ parameters (and as many corresponding invariants, see below) for the lepton sector, in agreement with~Ref.~\cite{Jenkins}. 

Note that the low-energy observable $m_\nu$, Eq.~(\ref{numass2}), contains $9$ parameters only ($4$ for the $U_{\rm PMNS}$ matrix, $3$ mass eigenvalues and $2$ Majorana phases). This is because we can factorize from ${\it Y}_\nu$ a complex orthogonal hermitian matrix, hence $3$ parameters,  which would drop from the 
expression in Eq.~(\ref{numass}),  see Ref.~\cite{Casas:2001sr}.

\section{Natural extrema of an invariant potential} 
We summarize here the elements to identify the {\it natural} extrema of an invariant potential $V(x)$, that is 
those extrema  that are less or not at all dependent from specific tuning of the coefficients in the potential, compared to the generic extrema. 
We do not make any assumption about the convergence of the expansion of the potential in powers of higher-dimensional invariants, as done e.g.~in
Ref.~\cite{Alonso:2012fy,Alonso:2013mca}.

\begin{figure}[t]
\begin{center}
\includegraphics[scale=.50]{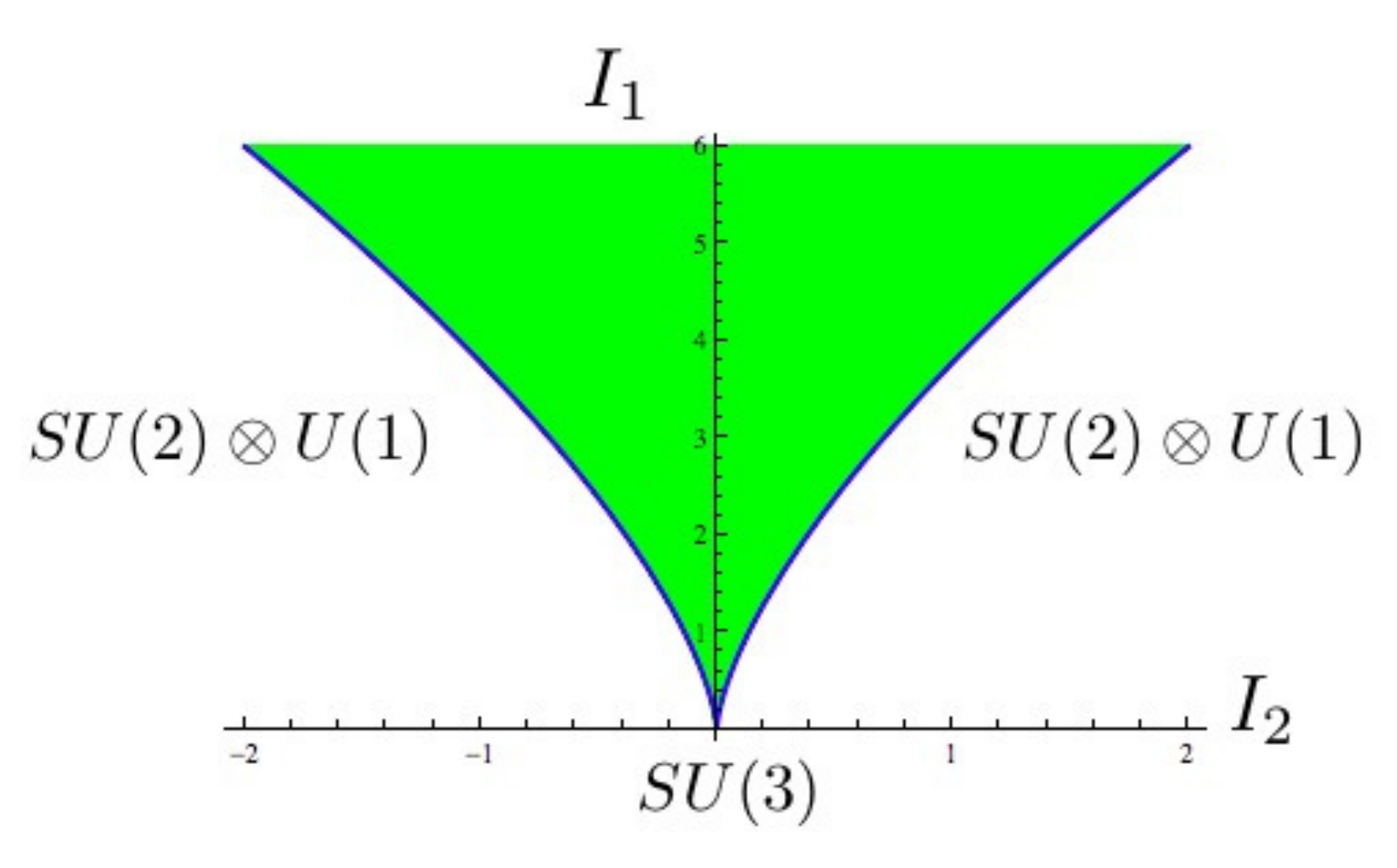}       
\caption{\label{fig:manifold}  {\footnotesize  Manifold ${\cal M}$ of the $SU(3)$ invariants constructed from $x$=octet=hermitian, $3\times 3$, traceless matrix (green region). Each point of ${\cal M}$ represents the orbit of  $x$, namely the set of points in octet space given by: $x_g=gxg^{-1}$, when $g$ runs over $SU(3)$. Boundaries of ${\cal M}$ are represented by Eq.~(\ref{boundary}). The little groups of the elements of different boundaries are indicated.  
}}
\end{center}
\end{figure}

The variables $x$ are the field components, transforming as given representations of the invariance group ${\cal G}$. 
In order to be invariant, $V(x)=V[I_i(x)]$, where $I_i$ are the independent invariants one can construct out of $x$. The crucial point is that the space of the $x$ has no boundary, while the manifold ${\cal M}$, spanned by $ I_i(x)$, does have boundaries. The situation is exemplified in Fig.~\ref{fig:manifold}, with ${\cal G}=SU(3)$, and $x$=octet=hermitian, $3\times 3$, traceless matrix. Defining the invariants $I_1={\rm Tr}(x^2)$ and $I_2={\rm Det}(x)$, 
the boundary is 
\be
I_1 \geq (54~I_2^2)^{1/3}~, \qquad-\infty < I_2 < +\infty~.
\label{boundary}
\ee

In general, let $N$ be the number of algebraically independent invariants.
One sees easily that~\cite{Michel:1970mua,Cabibbo:1970rza}:
\begin{itemize}
\item each point of ${\cal M}$ represents the orbit of  $x$, namely the set of points in octet space given by: $x_g=gxg^{-1}$, when $g$ runs over ${\cal G}$;
\item points on each boundary admit little (i.e. invariance) groups , which are the same up to a ${\cal G}$ conjugation.
\end{itemize}

%@@@@@@@@@@@@@@@@@@@@@@@@
%@@@@@@@@@@@@@@@@@@@@@@

 The boundaries of $\cal M$ are characterized by the rank of the Jacobian matrix being less than maximum~\cite{Cabibbo:1970rza}:
\be
J=\frac{\partial(I_1,I_2,\cdots)}{\partial(x_1,x_2,\cdots)}~, \qquad 
{\rm Rank}(J)=R<N~.
\ee
Boundaries are described by $R=N-1$ manifolds (e.g. surfaces, for $N=3$), each characterized by a different little group. Such manifolds  
 meet along $R=N-2$ dimensional manifolds (e.g. lines) which in turn meet along even lower dimensionality manifolds (e.g. singular points), etc..  
Each of these boundaries corresponds to a particular little group. 

The extrema of $V(x)$ are to be found by solving the equations:
\be
\frac{\partial V}{\partial x_j}=\sum_i \frac{\partial V}{\partial I_i}\frac{\partial I_i}{\partial x_j}=\sum_i \frac{\partial V}{\partial I_i} J_{ij}=0~.
\ee
We may now state the following.
\begin{itemize}
\item  $V$ has always extrema on boundaries having as little group a {\it maximal subgroup}\footnote{I.e. a subgroup that can be included  only in the full group $\cal G$.} of ${\cal G}$~\cite{Michel:1970mua};
\item extrema of $V$ with respect to the points of a given boundary are extrema of $V(x)$~\cite{Cabibbo:1970rza}.
\end{itemize}
The latter extrema are more natural than the generic extrema in the interior of $\cal M$, since they require the vanishing of only $N-1$, or $N-2$, etc. derivatives of $V$ given that, on the boundary, $J$ has 1 or 2, etc. vanishing eigenvectors (orthogonal to the boundary). Thus, from Fig.~\ref{fig:manifold} we learn that it is more natural to break $SU(3)$ along the direction of the hypercharge~\cite{Michel:1970mua} ($x$ with two equal eigenvalues, little group $SU(2)\otimes U(1)$) than along the direction of $T_3$, which corresponds to elements in the interior of $\cal M$.

The case of chiral $SU(3)\otimes SU(3)$ was considered in Ref.~\cite{Cabibbo:1970rza}, with the symmetry broken by quark masses in the $(3,{\bar 3})\oplus ({\bar 3},3)$. The elements of such representation are  complex, $3\times 3$, matrices, $M$, transforming according to:
\bea
&&M \to M^\prime=UMV^\dagger~.  \notag
\eea
By one such transformation, one can reduce $M$ to the standard diagonal, positive, form ( up to an irrelevant overall phase~\cite{Cabibbo:1970rza}):
\bea
&& M= {\bar U}m{\bar V}^\dagger~, \qquad m_q={\rm diag}(a,b,c)~.
\label{33bar}
\eea
There are three invariants, corresponding to the mass eigenvalues in Eq.~(\ref{33bar}):
\be
I_1={\rm Tr}(M M^\dagger)~, \qquad I_2= {\rm Tr}(M M^\dagger M M^\dagger)~, \qquad I_3= {\rm Det}(M)~,
\ee
and the natural extrema located on the boundaries correspond to the subgroups:
\be
 \begin{array}{lll}
 SU(2)\otimes U(1) & m_q={\rm diag}(a,a,c) & ({\rm exact~isospin},~ {\rm Rank=2})~,  \\
 SU(2)\otimes SU(2)\otimes U(1) \ & m_q={\rm diag}(0,0,c) & ({\rm hierarchical,~ Rank}=1,~{\rm maximal})~, \\
 SU(3) & m_q={\rm diag}(a,a,a) & ({\rm degenerate,~ Rank}=1,~{\rm maximal})~,  \\
SU(3)\otimes SU(3) & m_q={\rm diag}(0,0,0) & ({\rm symmetry~unbroken,~ Rank}=0)~.
\end{array}
\label{chiral}	
\ee

Turning to our case, we note that there are no common matrices in the transformations of quarks, Eq.~(\ref{quarktransf})  or leptons, Eq.~(\ref{leptransf}). 
This implies that the invariants\footnote{For an earlier analysis of invariants in view of minimizing flavor potentials, see Ref.~\cite{Jenkins:2009dy}.} divide in two independent sets: the Jacobian factorizes in two determinants and we can discuss the two cases separately.

\section{Quarks in three families} 
To classify the invariants, we define two matrices which transform in the same way under $SU(3)_q$ and are singlet under the
other transformations: 
\be
\rho_U=Y_U Y_U^\dagger~, \qquad \rho_D=Y_D Y_D^\dagger~;  \qquad   \rho_{U,D} \to U_{q}  \rho_{U,D} U_{q}^\dagger~.
\ee
There are six unmixed invariants, which we may take as:
\be
I_{U^{1}}={\rm Tr}(Y_U Y_U^\dagger)~, \quad   I_{U^{2}}={\rm Tr}[(Y_U Y_U^\dagger)^2]~, \quad I_{U^{3}}={\rm Tr}[(Y_U Y_U^\dagger)^3]~,
\ee
and the same for $Y_D Y_D^\dagger$. Next we define four mixed invariants:
\be
\begin{array}{ll}
 I_{U,D}={\rm Tr}( Y_U Y_U^\dagger Y_D Y_D^\dagger)~, \quad 
& I_{U^2,D}={\rm Tr}[(Y_U Y_U^\dagger)^2 Y_D Y_D^\dagger)~,   \\
 I_{U,D^2}={\rm Tr}[Y_U Y_U^\dagger (Y_D Y_D^\dagger)^2]~, \quad 
& I_{(UD)^2}={\rm Tr}[( Y_U Y_U^\dagger Y_D Y_D^\dagger)^2]~. 
\end{array}\label{mixQ}
\ee
 As anticipated, 10 independent invariants suffice to characterize in generality the physical degrees of freedom in
the Yukawa fields. We stress  in particular that the 4 invariants in Eq.~(\ref{mixQ}) contain enough information to reconstruct the 4 physical parameters of
the CKM matrix, including its CP-violating phase (up to discreet choices, see Ref.~\cite{Jenkins:2009dy}), despite none of them 
vanishes in the limit of exact CP invariance.

Unmixed invariants produce extrema corresponding to degenerate or hierarchical patterns as in the chiral case illustrated in (\ref{chiral}).
Mixed invariants involve the CKM matrix $U$, e.g.:
\be
I_{U,D}={\rm Tr}(Y_U Y_U^\dagger Y_D Y_D^\dagger)= \sum_{ij}U_{ij}U^\star_{ij}(m^2_U)_i(m^2_D)_j=\sum_{ij}P_{ij}(m^2_U)_i(m^2_D)_j~.
\ee

The matrix $P_{ij}=|U_{ij}|^2$  enjoys the properties that: all elements are between zero and 1, the sum of the elements of any row equals the sum of the elements of any column and both sums are equal to unity. Such matrices (by the so-called Birkhoff-Von Neumann theorem~\cite{vonneumann}) are convex combinations of permutation matrices, i.e.~matrices with a $1$ and all other null elements in each row, the $1$ being in different columns. Thus, permutation matrices provide us the singular points on the boundary of the domain, without having to compute the rank of the determinant. The upshot is that, after a relabeling of the $down$ quark coupled to each $up$ quark, we end up with $U_{\rm CKM} =1$.

A more detailed analysis is given in Ref.~\cite{rodrigothesis}. It involves the calculation of the $10\times 10$ Jacobian, which factorizes in two, $3\times 3$ Jacobians (unmixed invariants) and a $4\times 4$ one (mixed invariants) and it confirms the conclusion that one natural solution, for the three families quark case, is a hierarchical one, with dominating third family masses, and trivial  CKM matrix.

In the limit of vanishing masses for the first two generations, this solution corresponds to the little  group $SU(2)_q \otimes SU(2)_U \otimes SU(2)_D \otimes U(1)$
that is a maximal subgroup of $SU(3)_q \otimes SU(3)_U \otimes SU(3)_D$.

\section{Leptons in three families}
For leptons, we need 15 invariants. To construct them, we consider first the two combinations:
\be
\rho_E=Y_E Y_E^\dagger~, \qquad  \rho_\nu= Y_\nu Y_\nu^\dagger~;  \qquad   \rho_{E,\nu} \to U_{\ell}  \rho_{E,\nu} U_{\ell}^\dagger~, 
\ee
in which ${\cal O}(3)$ transformations disappear. We may construct unmixed and mixed invariants, as in the quark case, the mixed ones involving the matrix $U_L$, Eq.~(\ref{standardlep}). We choose the unmixed ones as:
\be
{\rm Unmixed, E:} \qquad 
I_{E^1}={\rm Tr}( Y_E Y_E^\dagger)~, \quad 
I_{E^2}={\rm Tr}[(Y_E Y_E^\dagger)^2]~, \quad 
I_{E^3}= {\rm Tr}[(Y_E Y_E^\dagger)^3]~,
\label{unmixE}
\ee
and three similar ones  ($I_{\nu^{1-3}}$) using $\rho_\nu$,  while the four mixed invariants containing $\rho_e$ and $\rho_\nu$
are taken to be:
\be
{\rm Mixed, ~type ~1}: \qquad 
\begin{array}{ll}
 I_{\nu,E}={\rm Tr}( Y_\nu Y_\nu^\dagger Y_E Y_E^\dagger )~, \quad  
 &  I_{\nu^2,E}={\rm Tr} [( Y_\nu Y_\nu^\dagger)^2 Y_E Y_E^\dagger ]~,  \\
 I_{\nu,E^2}={\rm Tr} [Y_\nu Y_\nu^\dagger (Y_E Y_E^\dagger)^2 ]  ~, \quad 
 & I_{(\nu E)^2}={\rm Tr}[( Y_\nu Y_\nu^\dagger Y_E Y_E^\dagger )^2] ~.
 \end{array}
\label{mixedenu}
\ee

For neutrinos we may construct also a matrix which transforms under the orthogonal group only:
\be
\sigma_\nu= Y_\nu^\dagger Y_\nu~;   \qquad   \sigma_\nu  \to  \cO  \sigma_\nu \cO^T~.
\ee
The symmetric and antisymmetric parts of $\sigma_\nu$ transform separately and can be used to construct two different invariants, such as 
${\rm Tr}[   Y_\nu^\dagger Y_\nu  (  Y_\nu^\dagger Y_\nu  \pm  Y_\nu^T  Y_\nu^* )]$. 
Here the first term in the product gives back the invariant  $I_{\nu^2}={\rm Tr} [( Y_\nu Y_\nu^\dagger)^2]$, but the second one gives rise to new contractions which involve the unitary, symmetric matrix 
\be
W=U_R U_R^T~.
\label{blocco}
\ee
We thus define the following three additional invariants:
\be
{\rm Mixed, ~type ~2}: \qquad 
\begin{array}{ll}
J_{\sigma^1}={\rm Tr}(Y_\nu^\dagger Y_\nu Y_\nu^T  Y_\nu^* )~, \quad 
& J_{\sigma^2}={\rm Tr}[(Y_\nu^\dagger Y_\nu)^2 Y_\nu^T  Y_\nu^* ]~, \\ 
 J_{\sigma^3}={\rm Tr}[(Y_\nu^\dagger Y_\nu Y_\nu^T  Y_\nu^*)^2]~. & 
\end{array}
 \label{LRnu}
 \ee
 Finally, we add two invariants which contain both $U_L$ and $W$:
 \be
{\rm Mixed, ~type ~3}: \qquad 
\begin{array}{ll}
I_{LR}=\mbox{Tr}\left[{\it Y}_\nu {\it Y}_\nu^T {\it Y}_\nu^* {\it Y}_\nu^\dagger {\it Y}_E {\it Y}_E^\dagger \right]~, \\
I_{RL}=\mbox{Tr}\left[{\it Y}_\nu {\it Y}_\nu^T{\it Y}_E^* {\it Y}_E^T  {\it Y}_\nu^* {\it Y}_\nu^\dagger {\it Y}_E {\it Y}_E^\dagger \right]~.
\end{array}
\qquad\qquad 
\ee

The discussion of the Jacobian leads to the following results, see again Ref.~\cite{rodrigothesis} for details.
\begin{itemize}
 \item Unmixed invariants produce extrema corresponding to degenerate or hierarchical mass patterns.
  \item Mixed, type 1, invariants contain $|(U_L)_{ij}|^2$ and lead, like in the quark case, to the conclusion that $U_L$ is a permutation matrix 
  {(up to an overall phase)}.
\item Mixed, type 2, invariants contain  $|W_{ij}|^2$ and indicate that $W=U_R U_R^T$ is a also permutation matrix  {(up to an overall phase)}.
\item {Once we impose that $U_L$ and $W$ are permutation matrices, the sensitivity of Mixed, type 3 invariants to $\omega$ vanishes. 
The latter remains therefore undetermined. }
 \end{itemize}
We may absorb the first permutation matrix in a relabeling of the neutrinos coupled to each charged lepton, 
but the second matrix leads to a non trivial result for the neutrino mass matrix, Eq.~(\ref{numass2}). The reason for the difference is that, for quarks we could eliminate any complex matrix $U_D$ by a redefinition of $D_R$, but this is not possible for leptons, because we can redefine the $N_i$ only with a real orthogonal matrix.

We use the freedom in the neutrino labeling to set $U_L=1$ in the basis where charged leptons are ordered according to: 
\be
{\it Y}_E=\mbox{diag}\,(y_e, y_\mu,y_\tau)~.
\label{chlept}
\ee
There are four possible symmetric permutation matrices that can be associated with $W=U_R U_R^T$, one of them being the unit matrix. The other three imply non trivial mixing in one of the three possible neutrino pairs, e.g.
{
\be
W=U_R U_R^T=-\left( \begin{array}{ccc} 1&0&0\\0&0&1\\0&1&0 \end{array}\right)
\ee
We introduced the minus sign for $W$ to have a positive determinant,\footnote{{We thank  E. Nardi for useful discussions about this point and the role of $\omega$ in Eq.~(\ref{numass2}).}} consistently with the  condition ${\rm Det}(U_R)=1$. 

Using this expression in Eq.(\ref{numass2}) leads to 
\be
m_\nu=\frac{v^2}{M}~y_\nu \omega  W  \omega  y_\nu=\frac{v^2}{M}~\left(\begin{array}{ccc}-y_1^2 e^{2i\alpha} & 0 & 0\\ 0 & 0 & -
y_2 y_3 e^{-i\alpha}  \\0 &- y_2 y_3 e^{-i\alpha}  & 0 \end{array}\right)~,
\label{solution}
\ee
where $y_\nu=$~diag($y_1,y_2,y_3$) and $\omega$= diag($e^{i\alpha} , e^{i\beta} , e^{-i(\alpha+\beta}$).  
The absence of mixing between the first eigenvector of $m_\nu$ and those associated to the 2-3 sector implies that the 
phase $\alpha$ is unphysical and can be set to zero by an appropriate phase redefinition of the neutrino fields.
From the second identity in Eq.~(\ref{numass2}) we then find:
\bea
&& {\hat m}_\nu=\frac{v^2}{M}~{\rm diag}(y_1^2,y_2y_3,y_2y_3)~,\notag \\
&&U_{\rm PMNS}^{(0)}=\left(\begin{array}{ccc}1 & 0 & 0\\ 0 & 1/\sqrt{2} & 1/\sqrt{2}\\ 0 & -1/\sqrt{2} & 1/\sqrt{2} \end{array}\right)~, 
\qquad \Omega= {\rm diag}(-i,-i,1)~.
\label{pmns0}
\eea
The non-trivial relative Majorana phase in the 2-3 sector}
 is needed to bring all masses in positive form. The one maximal mixing angle and one maximal  Majorana phase stem from the $\cO(2)$ substructure in Eq.~(\ref{solution}), as found in Ref.~\cite{Alonso:2012fy,Alonso:2013mca}. 

With three families we can go closer to the physical reality if we assume complete degeneracy for $y_\nu$. In this case, after the $2-3$ rotation we are left with degenerate $1$ and $2$ neutrinos and, a priori, a new rotation will be needed to align the neutrino basis with the basis in which the charged lepton mass takes the diagonal form in Eq.~(\ref{chlept}). We may expect, in this case, the PMNS matrix to have an additional rotation in the $1-2$ plane:
\be
U_{\rm PMNS}=U_{\rm PMNS}^{(0)} ~U(\theta_{12})~.
\ee
We shall see that small perturbations around the solution in Eq.~(\ref{solution}) allow to determine this angle, that remains non-zero 
in the limit of vanishing perturbations.

\section{ Group theoretical considerations} 
One may ask what is the little group corresponding to the extremal solution, Eq.~(\ref{solution}). While ${\it Y}_\nu$ transforms under $SU(3)_{\ell}\otimes {\cal O}(3)$, orthogonal transformations drop out of ${\it Y}_\nu{\it Y}_\nu^T$. In some sense we have to find the appropriate square root of $m_\nu$.  
By explicit calculation, one sees that the answer is given by\footnote{${\it Y}_\nu$ is uniquely determined up to an inessential right multiplication by an orthogonal matrix.}:
\bea
&&Y_\nu=
{
\left(\begin{array}{ccc}i y_1 & 0 & 0 \\ 0 &i \frac{y_2}{\sqrt{2}} &  \frac{y_2}{\sqrt{2}} \\ 0 & i \frac{ y_3}{\sqrt{2}} &-\frac{y_3}{\sqrt{2}} \end{array}\right)~.
}
\label{squareroot}
\eea
${\it Y_\nu}$ transforms under $SU(3)_{\ell}\otimes {\cal O}(3)$ according to the $(\bar 3,3_V)$ representation, where the suffix V denotes the vector representaton of $\mathcal O (3)$, realized, in triplet space, by the Gell-Mann imaginary matrices $\lambda_{2,5,7}$. One verifies that:
\be
\lambda_3^\prime Y_\nu-Y_\nu \lambda_7=0;~\lambda_3^\prime={\rm diag}(0,1,-1)~,
\ee
i.e.~for this solution, $SU(3)_{\ell}\otimes {\cal O}(3)$ is reduced to the $U(1)_{\rm diag}$ subgroup of transformations of the form:
\be
U(1)_{\rm diag}:~{\rm exp} \left(i\epsilon \lambda_3^\prime\right)\otimes {\rm exp}\left(i\epsilon\lambda_7\right)~.
\ee
 This $U(1)_{\rm diag}$ is the little group of the boundary to which the solution in Eq.~(\ref{squareroot}) belongs. 
When combined with a hierarchical solution for the charged-lepton Yukawa of the type $Y_E\propto (0,0,1)$,
this corresponds to the little group $SU(2)_E \otimes U(1)_{\rm diag}$,
a subgroup of $SU(3)_E \otimes SU(3)_\ell \otimes {\mathcal O(3)}$.

In the limit  $y_1= y_2 =y_3$, ${\it Y}_\nu$ becomes proportional to a unitary matrix:
\bea
&&Y_\nu \to y~
{
\left(\begin{array}{ccc}i & 0 & 0 \\ 0 & i \frac{1}{\sqrt{2}} & \frac{1}{\sqrt{2}} \\ 0 & i \frac{1}{\sqrt{2}} &- \frac{1}{\sqrt{2}} \end{array}\right)
}
=y V~,
\qquad V V^\dagger =1~,
\label{eq:40}
\eea
and the $U(1)$ invariance is augmented to a full ${\cal O}(3)_{\rm diag}$, a maximal subgroup of $SU(3)_{\ell}\otimes {\cal O}(3)$:
\be
Y_\nu\to (V{\cal O}V^\dagger) Y_\nu {\cal O}^T= Y_\nu~,
\ee 
where ${\cal O}$ is an orthogonal matrix  generated by $\lambda_{2,5,7}$. The ${\cal O}(3)_{\rm diag}$ would remain 
unbroken only in the case of degenerate charged lepton masses. Combining $Y_\nu$ in  Eq.~(\ref{eq:40})
 with $Y_E\propto (0,0,1)$,
we recover the little group $SU(2)_E\otimes U(1)_{\rm diag}$. 

Summarizing:
\begin{itemize}
\item{} $Y_E\propto (0,0,1): \quad SU(3)_{E} \otimes SU(3)_\ell \to SU(2)_E\otimes SU(2)_\ell \otimes U(1) \quad$ (maximal~subgroup) 
\item{} $Y_\nu$ in  (\ref{squareroot})~:  $\qquad\quad  {\hat m}_\nu={\rm diag}(m_1,m,m)~,
\qquad~SU(3)_{\ell}\otimes {\cal O}(3)\to U(1)_{\rm diag}$
\item{} $Y_\nu$ in (\ref{eq:40})~: $\quad\quad {\hat m}_\nu=m\times 1~, \qquad  ~ SU(3)_{\ell}\otimes {\cal O}(3) \to {\cal O}_{\rm diag}(3)$
\quad  (maximal subgroup)
\item{} $Y_E\propto (0,0,1)$ \& $Y_\nu$ in (\ref{squareroot}) or (\ref{eq:40})~:
  $\quad\ SU(3)_{E} \otimes SU(3)_{\ell}\otimes {\cal O}(3) \to  SU(2)_E\otimes U(1)_{\rm diag}$
\end{itemize}
Both breaking patterns of $Y_\nu$ feature: i) at least two degenerate neutrinos; ii)~$\theta_{23}=\frac{\pi}{4}$ and $\theta_{13}=0$;
iii)  one {maximal Majorana phase}. In addition, the degenerate pattern in Eq.~(\ref{eq:40})  implies 
three degenerate neutrinos and a second large (not calculable) mixing angle.

We finally note that all the features of the degenerate pattern can also be obtained without assuming the existence of  heavy  
right-handed neutrinos, but rather starting from the flavor symmetry group ${\cal G}_0$ in Eq.~(\ref{gnumassless}) and assuming that the effective (light) neutrino mass matrix, 
\be
\frac{1}{v^2 }~{\bar {\ell }}_L  m_\nu  {\tilde H} {\tilde H}^T  {\ell}^C_L~, 
\ee
breaks $SU(3)_\ell$ into the maximal subgroup $\mathcal O (3)_\ell$. This breaking pattern implies a neutrino mass matrix proportional to a unitary matrix.
The latter must be a symmetric permutation matrix (in the basis where $Y_E$ is diagonal) in order to leave an unbroken $ U(1)$
when combined with  the charged-lepton Yukawa coupling. Selecting $U(1)_{\rm diag}$ among the possible unbroken subgroups 
we recover $m_\nu$ in Eq.~(\ref{pmns0}) with $y_1^2=y_2y_3$.

\section{Small perturbations}
We now consider the addition of small perturbations to the matrix in Eq.~(\ref{pmns0}),  
see for instance Refs.~\cite{Altarelli:2004za},~\cite{Blankenburg:2012nx}.  
For simplicity, we analyze in detail the case of real perturbations. Under this assumption,
the most general form of the perturbation is: 
\be
m_\nu ={-} \frac{v^2 y}{M}\left(\begin{array}{ccc}1+\delta +\sigma & \epsilon+\eta & \epsilon-\eta\\\epsilon+\eta & \delta +\kappa & 1\\\epsilon-\eta & 1& \delta-\kappa \end{array}\right)~,
\ee
with $|\epsilon,\eta,\sigma,\delta,\kappa|\ll1$. 
A simple calculation leads to the first order results:
\bea
&&{\hat m}_\nu = m_0~{\rm diag}\left(1+\delta+\sqrt{2}\epsilon \frac{c}{s}  , 1+\delta-\sqrt{2}\epsilon \frac{s}{c}  , 1-\delta \right)~, \label{massdiag} \\
&&{\Omega={\rm diag}(-i,-i,1)}~, \label{eq:Majo} \\
&& U_{\rm PMNS}=\left[ \begin{array}{ccc}
c  & -s & \displaystyle\frac{\eta}{\sqrt{2}} \\
               \displaystyle{\frac{s}{\sqrt{2}}}\left(1 + \frac{c}{s} \frac{\eta}{\sqrt{2}} +\frac{\kappa}{2}\right) \
           &  \displaystyle{\frac{c}{\sqrt{2}}}\left( 1- \frac{s}{c}\frac{\eta}{\sqrt{2}}+\frac{\kappa}{2}\right) \
           & -\displaystyle\frac{1}{\sqrt{2}} \left(1-\frac{\kappa}{2}\right)\\ 
               \displaystyle{\frac{s}{\sqrt{2}}}\left(1 - \frac{c}{s} \frac{\eta}{\sqrt{2}} -\frac{\kappa}{2}\right) \
           &  \displaystyle{\frac{c}{\sqrt{2}}}\left( 1+ \frac{s}{c}\frac{\eta}{\sqrt{2}}-\frac{\kappa}{2}\right) \
           & +\displaystyle\frac{1}{\sqrt{2}} \left(1+\frac{\kappa}{2}\right) \end{array}\right]~,
\eea
where $s=\sin(\theta_{12})$, $c=\cos(\theta_{12})$, and $\tan(2\theta_{12})=2\sqrt{2}\epsilon/\sigma$.

The PMNS matrix features a generically large $\theta_{12}$ (that we cannot compute in absence of firm predictions for the values of 
$ \epsilon$ and $\sigma$, but that does not goes to zero in the limit of vanishing perturbations), 
$\theta_{23}$ close to $\pi/4$, and $\theta_{13}$ generically small. With a  suitable choice of the perturbations one can 
easily achieve the so-called bimaximal or tribimaximal mixing form.

The spectrum is almost degenerate, with normal or inverted hierarchy according to the signs of the perturbations, and 
mass splittings not correlated to the mixing matrix. Determining the size of the perturbations from 
$|\sin\theta_{13}|$ or, equivalently,  from the deviation of $\theta_{23}$ from $\pi/4$, and assuming a similar size for the 
perturbation controlling the largest mass splitting, we find 
\be 
\frac{|\Delta m^2_{\rm atm}|}{2 m^2_0} \approx| \sin\theta_{13}|\approx |\theta_{12}-\frac{\pi}{4}|\approx 0.1 \quad 
\longrightarrow  \quad  m_0 \approx 0.1~{\rm eV}~.
\label{eq:m0}
\ee
A lightest neutrino mass of this size is within reach of the next generation of $0\nu\nu$ double beta decay experiments~\cite{pdg},  and possibly of
cosmological measurements \cite{Ade:2013zuv,Ade:2013lmv}.
Note also that the size of the perturbations is not far from what could be deduced from the charged lepton spectrum, 
treating $m_\mu/m_\tau \approx 0.06$ as estimate of the sub-leading terms.

So far we considered only real perturbations. Assuming complex perturbations does not change qualitatively the results 
listed above for the mass spectrum and the mixing angles, but leads to non-vanishing CP-violating phases. In particular, 
we find a generically large Dirac-type phase {and a generically large second 
Majorana phase (that becomes physical), while small corrections to the maximal Majorana phase in the 2-3 sector.} 
In this context, we note that small Majorana phases are enough for successful leptogenesis with almost degenerate right-handed neutrinos
masses (see e.g.~Ref.~\cite{O3Rleptogenesis}).

In this letter  we shall not try  to identify the origin of the small perturbations. Given the non-perturbative character of the argument that leads to the $U(1)_{\rm diag}$ symmetric boundary, the latter symmetry cannot obviously be lifted by ${\cal G}$ symmetric interactions in higher perturbative order. Small perturbations may result however from the effect of other fields, transforming differently from the $Y$'s and acquiring smaller vevs, like e.g.~in Refs.~\cite{Barbieri:1995uv, Alonso:2011yg, Blankenburg:2012nx}, or by the effect of interactions external to the present scheme, e.g.~arising from gravity.

\section{Conclusions and outlook} 
We have assumed that the structure of quark and lepton mass matrices derives
from a minimum principle, with the maximal flavor symmetry $\left[SU(3)\right]^5\otimes  {\mathcal O}(3)$
 and a minimal breaking due to the vevs of fields 
transforming like the Yukawa couplings.  For leptons we find a natural solution correlating large mixing angles and degenerate neutrinos. This solution 
generalizes to three familes and arbitrary invariant potential the results found in Ref.~\cite{Alonso:2012fy,Alonso:2013mca}.
The generalization of this result to arbitrary see-saw models has also been discussed. 
Subject to small perturbations, the solution can reproduce the observed pattern of neutrino masses 
and mixing angles. Our considerations lead to a value of the common neutrino mass 
that is within reach of the next generation of neutrinoless double beta decay experiments, and possibly within that of cosmological measurements.

\section*{Acknowledgments}
We thank D. Hernandez, L. Merlo and S. Rigolin for interesting discussions and comments on the preliminary version of this letter.
 We are also indebted to E. E. Jenkins, A. V. Manohar,  A. Melchiorri and S. Pascoli for useful discussions.  
 The authors acknowledge the stimulating environment and lively physics exchanges with the colleagues at the CERN Theory Group.  R.A.~and~M.B.~acknowledge partial support by the European Union FP7 ITN INVISIBLES (Marie Curie Actions, PITN-GA-2011-289442),  
as well as support from CiCYT through the project FPA2009-09017,  CAM through the project HEPHACOS P-ESP-00346, and from  European Union FP7 
ITN UNILHC (Marie Curie Actions, PITN-GA-2009-237920). R.A.~acknowledges MICINN support through the grant BES-2010-037869.
G.I.~acknowledges partial support by MIUR under project 2010YJ2NYW.

%@@@@@@@@@@@@@@@@@@@@@@@@@@@@@@@@@@@@@@@@@@@@@@

\end{document}